\documentstyle[11pt,newpasp,twoside,psfig]{article}

\def\edcomment#1{\iffalse\marginpar{\raggedright\sl#1\/}\else\relax\fi}
\reversemarginpar
\begin{document}

\title{THE SMALL SCALE STRUCTURE OF N103B - NATURE OR NURTURE?}
\author{KAREN T. LEWIS, DAVID N. BURROWS, GORDON P. GARMIRE, JOHN A. NOUSEK}
\affil{THE STATE UNIVERSITY OF PENNSYLVANIA, 525 DAVEY LABORATORY, UNIVERSITY PARK, PA, 16802, USA}

\author{JOHN P. HUGHES}
\affil{DEPARTMENT OF PHYSICS AND ASTRONOMY,RUTGERS UNIVERSITY,126 FRELINGHUYSEN ROAD, PISCATAWAY, NJ 08854, USA}

\author{PATRICK SLANE}
\affil{HARVARD-SMITHSONIAN CENTER FOR ASTROPHYSICS, 60 GARDEN STREET, MA, 02138, USA}
\begin{abstract}
We present new results from a 40.8 ks {\it Chandra} ACIS observation
of the young supernova remnant (SNR) N103B located in the Large
Magellanic Cloud. The high resolution {\it Chandra} image reveals
structure at the sub-arcsecond level, including several bright knots
and filaments. Narrow-band imaging and spatially resolved spectroscopy
reveal dramatic spectral variations in this remnant as well. In this
paper we discuss whether these variations are due to inhomogeneities
in the surrounding environment or were generated in the explosion which
created the SNR.
\end{abstract}
\section{Introduction}
\indent N103B, one of the brightest X-ray and radio sources in the Large
Magellanic Cloud (LMC), is a young, compact (D=7pc or 30$''$)
supernova remnant (SNR). The remnant lies on the north-eastern rim of
the H~II region N103 and is only 40 pc from the young star cluster NGC
1850 (Dickel \& Milne 1995). At radio and X-ray wavelengths, N103B has
a shell-like morphology with the western half being $\sim 3$ times
brighter than the eastern half (Fig 1). The H$\alpha$ image is
dominated by 4 bright knots in the west, but also shows a partial
shell as well (Williams et al. 1999). Due to its proximity to a star
forming region, it was originally suspected that N103B was the result
of the core collapse of a massive object. However, the ASCA spectrum
shows no evidence for K-shell emission from O, Ne or Mg while Si, S,
Ar, Ca, and Fe features are strong, indicating that the remnant is
more likely the result of a Type Ia SN explosion (Hughes et
al. 1995).\\
\begin{figure}
\centerline{\psfig{file=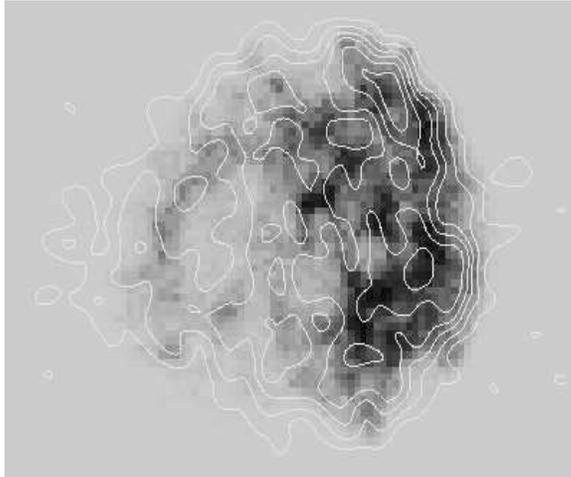,height=2.5in,width=3.0in}}
\caption{Plot of X-ray image with 2.5 cm ATCA contours (Dickel \& 
Milne 1995). Contours are 1.7, 1.3, 0.9, 0.6, 0.3, and 0.1
mJy/beam. The spatial resolution of the radio data is 1.2$''$, 
which is slightly larger than that of {\it Chandra}. }
\label{radio}
\end{figure}
\section{Observations \& Reduction}
\indent N103B was observed for 40.8 ks on 4 December 1999 using the
back-illuminated {\it Chandra} ACIS-S3 detector. The data were
corrected for Charge Transfer Inefficiency (CTI) as described in
Townsley et al. (2001a). The data were then cleaned and filtered with
the standard CIAO 2.1 software packaged. The ASCA grade filter
(g02346) was used throughout. The background was selected from an
annulus (r$_{in}=60''$, r$_{out}=120''$). We employ the response
matrices created by Townsley et al. (2001b) to match the CTI-corrected
data.
\section{Analysis \& Results}
\subsection{Brightness Variations}
\indent The most striking structural feature in N103B is the large 
brightness contrast between the eastern and western halves of the
remnant, which is present in the radio, X-ray, and, to some extent,
H$\alpha$ images. A cloud of absorbing material {\it could} produce
the observed brightness variation in X-ray. However, this same cloud
would not absorb the radio synchrotron emission. The radio and X-ray
brightness are generally very well correlated (Scronce et al., this
proceedings), thus it is much more likely that this brightness
variation is due to a density contrast between the eastern and western
halves of the remnant.\\
\indent In the high-resolution {\it Chandra} image, the western half 
of the remnant has been resolved into several filaments and bright
knots, which roughly co-incide with the H$\alpha$ knots. (The optical
image is not yet publicly available for direct comparison.) The
spectra of these knots do not show an enhancement of the heavy
elements, so it is unlikely that they are ``bullets'' of ejecta. One
possibility is that the H$\alpha$ knots, which show the normal LMC
abundances, (Russel \& Dopita 1992) are actually dense clumps in the
H~II region. As the dense cloudlets are heated by the shock, some of
the material will be evaporated and heated to X-ray emitting
temperatures, leading to en enhancement of the X-ray emission around
the dense H$\alpha$ clumps. Thus, many of the large brightness
variations, on both large and small spatial scales, are likely to be
caused by inhomogeneities in the surrounding environment.
\subsection{Narrow Band Imaging}
\indent We have extracted images of the remnant in the He-like Si, 
H-like Si, and S lines. The continuum was estimated by performing a
linear interpolation between two regions on either side of the
line. The Continuum Subtracted (CS) and Equivalent Width (EW) images
for He-like Si are shown in Fig. 2 ab. The CS image is quite similar
to the 0.5-8 keV image (Fig. 1). The EW image, however, reveals a
striking ring-like structure near the rim of the remnant. To ensure
that the ring-like structure is statistically significant, the image
was collapsed into the radial and azimuthal dimensions (Fig. 2cd).  As
can be seen, the EW of He-like Si increases radially with little
azimuthal dependence. A similar structure is seen in the H-like Si and
S EW images. The spectra of the innermost and outermost rings are
shown in Fig. 3; the increase in equivalent width of Si and S, as well
as Ar, is clear.
\begin{figure}
\centerline{\psfig{file=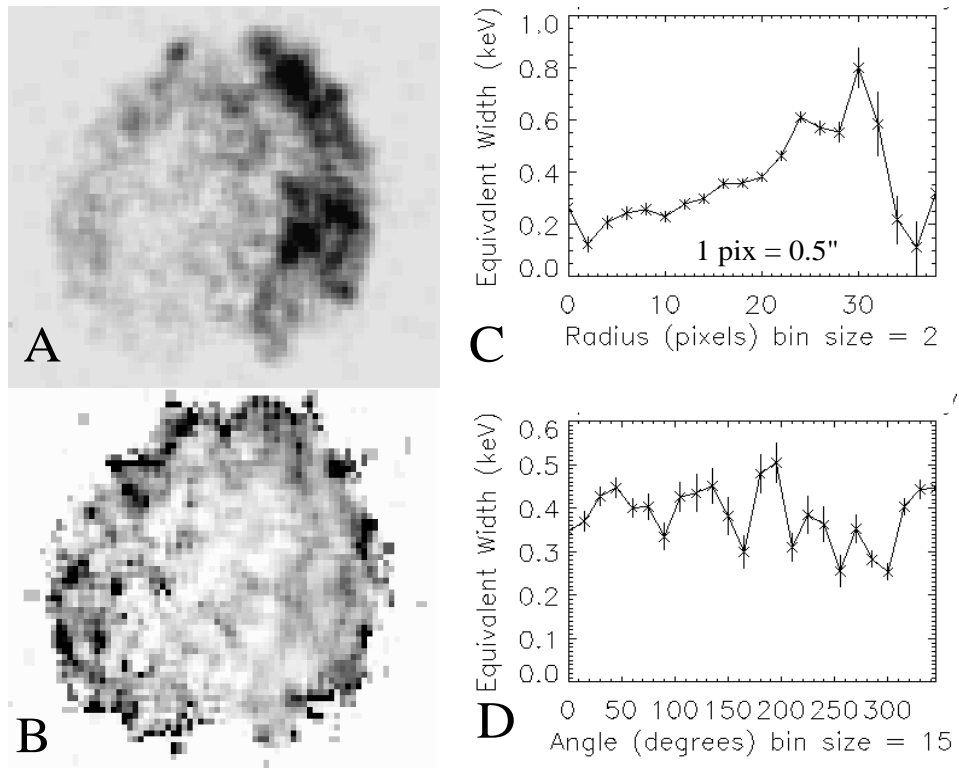,height=4in,width=5in}}
\caption{Narrow Band imaging in He-like Si. On the left, the continuum 
subtracted (A) and equivalent width (B) images. On the left, the radial 
(C) and azimuthal (D) dependence of the equivalent width.}
\label{narrowband}
\end{figure}
\begin{figure}
\centerline{\psfig{file=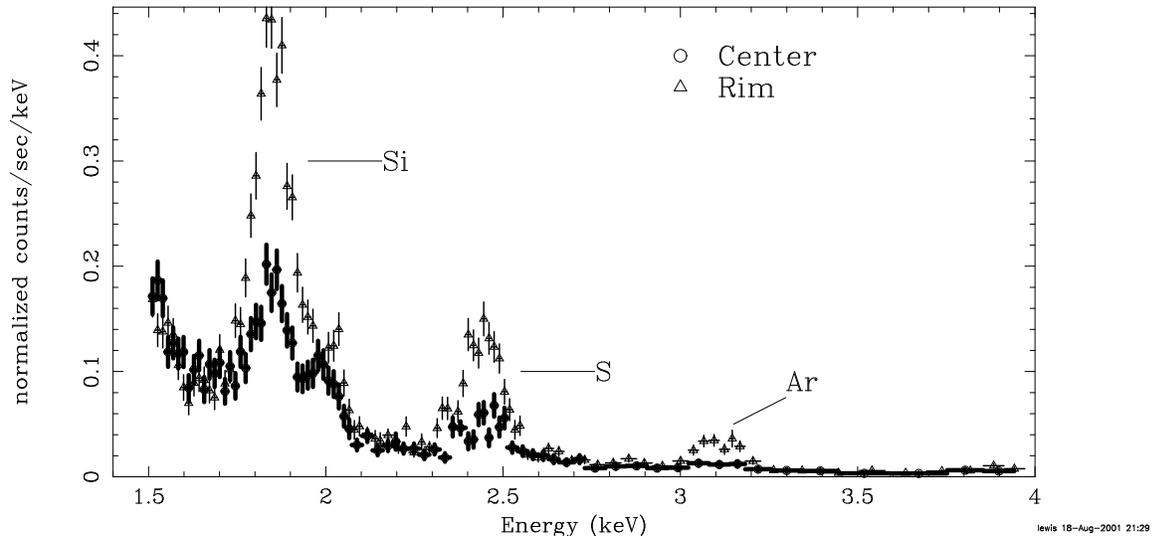,height=2.8in,width=6.0in}}
\caption{Comparison of the Si, S, and Ar line strengths in the 
inner and outermost rings.}
\label{specs}
\end{figure}

\subsection{Spectral Analysis}
\indent The increase in EW could be the result of temperature, 
ionization or abundance variations. To determine the true cause, we
divided the remnant into seven annuli, each with $\sim$ 35,000 counts,
and performed detailed spectral modeling on each.\\
\indent We use a non-equilibrium ionization (NEI) model developed 
by Jack Hughes along with two absorbing columns: one for the galaxy
and another for absorption within the within the LMC. The NEI model
allows for each element to have independent temperatures and
ionization states, which can be linked as desired. To reproduce the
spectra we find it necessary to have three thermodynamic ``phases'',
each of which contributes both to the line and continuum emission. The
predominant source of emission arises from a highly ionized
(n$_{e}$t$\ge 10^{12.5}$ s/cm$^{3}$), 0.9 keV plasma containing H, Si,
S, and Ar. To reconcile the emission from the L-shell transitions of
Fe and Ca with the strong K$\alpha$ line of these elements, it is
necessary to place Fe and Ca (as well as H) in a very high temperature
($\ge 3$keV), low ionization (n$_{e}$t$\sim$10$^{10-11}$ s/cm$^{3}$)
plasma. It is reasonable for Fe to be in a high temperature, low
ionization state since radioactive decay of Ni into Fe will locally
heat the Fe clumps, forming hot, diffuse ``bubbles'' of Fe. A similar
effect has been seen in Tycho's Remnant, another type Ia SNR (Hwang et
al. 1998). Finally, the O, Ne, and Mg emission is not consistent with
the high ionization state of the Si, S, and Ar. Thus we have a third
plasma which contains H, O, Ne, and Mg with an ionization timescale of
n$_{e}$t$\sim$10$^{11}$ s/cm$^{3}$. The temperature is not well
constrained, but is consistent with 0.9 keV.  Finally, a Gaussian
component at 0.72 keV was required in all but the outermost rings. We
believe that there may be several Fe L-shell transitions that are
missing in the model.\\
\indent An example fit is shown in Fig. 4. We find that, within error 
bars, the temperatures and ionizations of the different phases do not
vary significantly from ring to ring. The parameters which change most
dramatically are the abundances of Si, S, and Ar (Fig. 5).  The
abundances of these elements are relatively flat in the interior, then
rise sharply at a radius of 10$''$, similar to what is seen in the
equivalent width plots (Fig. 2cd). Thus the increase in the equivalent
width of Si, S, and Ar is due to an increase in the abundance of these
elements as opposed to a temperature or ionization variation. This
radial stratification of the Si-burning products is most certainly a
product of the supernova explosion itself.\\
\begin{figure}
\centerline{\psfig{file=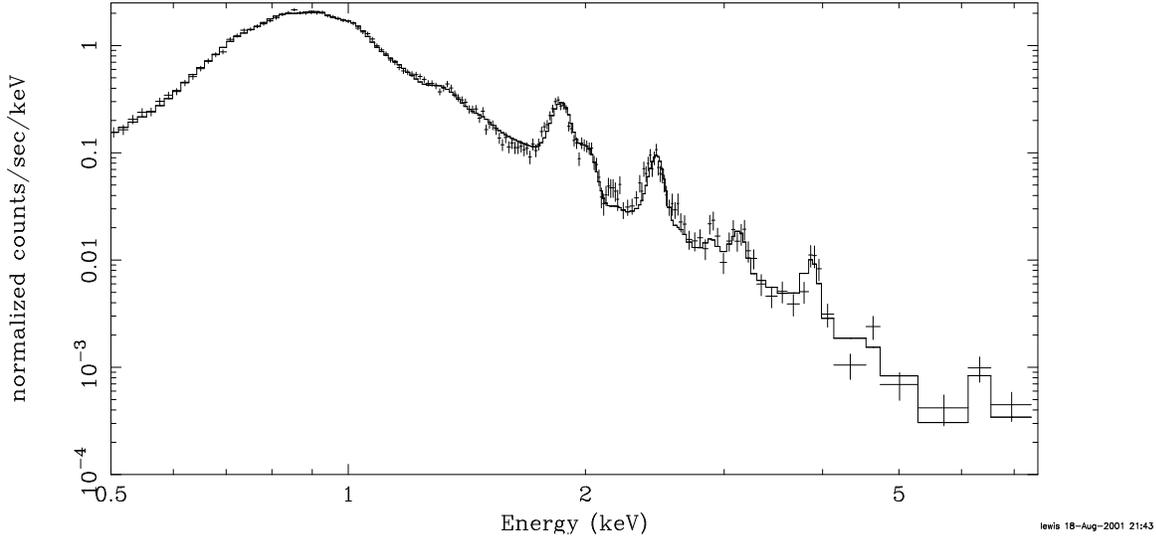,height=2.8in,width=6in}}
\caption{An example model fit for one of the rings. Note 
that the overall agreement between the data and model is
excellent. However, there are a few lines, particularly from the
K$\beta$ (n=3 to n=1) transitions, which are under-produced, 
indicating that the electron temperature may be higher than 
suggested by the model.}
\label{ring4}
\end{figure}
\begin{figure}
\centerline{\psfig{file=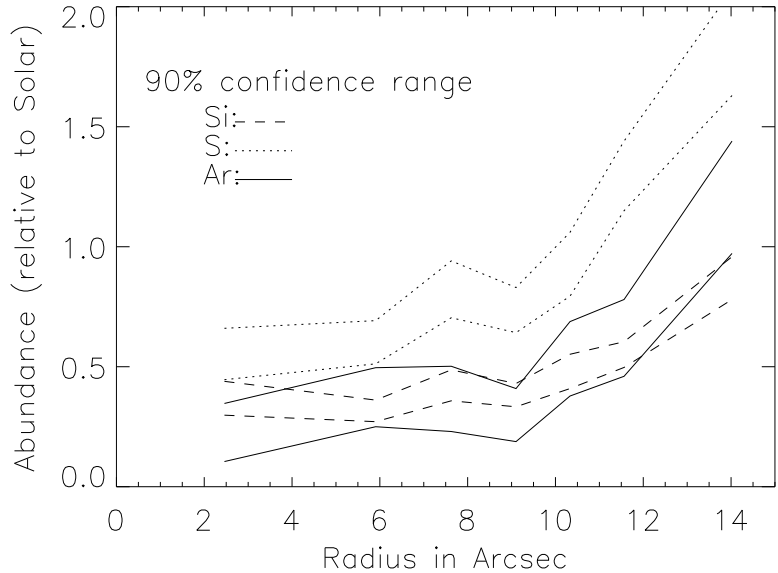,height=2.95in,width=4in}}
\caption{The 90 \% confidence intervals for the abundances 
of Si, S, and Ar. The increase is the abundances far exceeds the
random fluctuations, which are within the error bars.}
\label{abund}
\end{figure}
\section{Conclusions}
\indent This high resolution {\it Chandra} observation has revealed 
brightness variations on the sub-arcsecond level. We believe that many
of the variations are due to density inhomogeneities in the
surrounding medium. \\
\indent N103B appears to be very will mixed macroscopically and we see no
evidence for bullets of ejecta, as seen in many galactic remnants such
as Cas A (Hughes et al. 2000). This is partly due
to the fact that the physical resolution element is $\sim$10 times
larger in the LMC than within the galaxy. The plasma appears to possess
three distinct thermodynamic phases; on a microscopic level, N103B is
still quite clumpy. In particular we see evidence of a hot ($\ge
3$keV), low ionization (n$_{e}$t$\sim 10^{10-11}$s/cm${^3}$) Fe component
in the plasma such as that seen Tycho's SNR.\\
\indent We see an increase in the equivalent widths of Si, S, and Ar 
which are due to an increase in the abundances of these elements, as
opposed to a change in the temperature or ionization state. This
radial abundance stratification is most certainly a product of the
explosion itself, as opposed to a variation in the surrounding
medium. Future work will focus on the spectral analysis on the
non-radial variations in N103B, which should yield interesting
information on the interactions between this young supernova
remnant and its unusual environment.\\

\end{document}